\input harvmac
\def\AdS{{\rm AdS}}
\newcount\figno
\figno=0
\def\fig#1#2#3{
\par\begingroup\parindent=0pt\leftskip=1cm\rightskip=1cm\parindent=0pt
\global\advance\figno by 1
\midinsert
\epsfxsize=#3
\centerline{\epsfbox{#2}}
\vskip 12pt
{\bf Fig. \the\figno:} #1\par
\endinsert\endgroup\par
}
\def\figlabel#1{\xdef#1{\the\figno}}
\def\encadremath#1{\vbox{\hrule\hbox{\vrule\kern8pt\vbox{\kern8pt
\hbox{$\displaystyle #1$}\kern8pt}
\kern8pt\vrule}\hrule}}
\def\underarrow#1{\vbox{\ialign{##\crcr$\hfil\displaystyle
 {#1}\hfil$\crcr\noalign{\kern1pt\nointerlineskip}$\longrightarrow$\crcr}}}
%
\overfullrule=0pt

%

\def\bar{\overline}
\def\Z{{\bf Z}}

\def\R{{\bf R}}

\font\zfont = cmss10 
\font\litfont = cmr6

\def\bigone{\hbox{1\kern -.23em {\rm l}}}
\def\ZZ{\hbox{\zfont Z\kern-.4emZ}}
\def\half{{\litfont {1 \over 2}}}

\newcount\figno
\figno=0
\def\fig#1#2#3{
\par\begingroup\parindent=0pt\leftskip=1cm\rightskip=1cm\parindent=0pt
\global\advance\figno by 1
\midinsert
\epsfxsize=#3
\centerline{\epsfbox{#2}}
\vskip 12pt
{\bf Fig. \the\figno:} #1\par
\endinsert\endgroup\par
}
\def\figlabel#1{\xdef#1{\the\figno}}
\def\encadremath#1{\vbox{\hrule\hbox{\vrule\kern8pt\vbox{\kern8pt
\hbox{$\displaystyle #1$}\kern8pt}
\kern8pt\vrule}\hrule}}
\def\underarrow#1{\vbox{\ialign{##\crcr$\hfil\displaystyle
 {#1}\hfil$\crcr\noalign{\kern1pt\nointerlineskip}$\longrightarrow$\crcr}}}
%
\overfullrule=0pt

%

\def\bar{\overline}
\def\Z{{\bf Z}}

\def\R{{\bf R}}

\font\zfont = cmss10 
\font\litfont = cmr6

\def\bigone{\hbox{1\kern -.23em {\rm l}}}
\def\ZZ{\hbox{\zfont Z\kern-.4emZ}}
\def\half{{\litfont {1 \over 2}}}

\Title{hep-th/9910245}
{\vbox{\centerline{Connectedness Of The Boundary}
\bigskip
\centerline{ In The AdS/CFT Correspondence}}}
\smallskip
\centerline{Edward Witten}
\smallskip
\centerline{\it Department Of Physics, California Institute of Technology,
Pasadena, CA 91106}
\centerline{\it CIT-USC Center For Theoretical Physics}
\centerline{\it and}
\centerline{\it Institute For Advanced Study, Olden Lane, Princeton, NJ 08540}\bigskip
\centerline{and}
\bigskip
\centerline{S.-T. Yau}
\smallskip{\it Department of Mathematics, Harvard University, Cambridge
Mass. 02138}
\bigskip
\medskip

\noindent
Let $M$ be a complete Einstein manifold of negative curvature,
and assume that (as in the AdS/CFT correspondence)
it has a Penrose compactification with a conformal boundary
$N$ of positive scalar curvature.  We show that under these
conditions, $H_n(M;\Z)=0$ and in particular $N$ must be connected.
These results resolve some puzzles concerning the AdS/CFT correspondence.
\Date{October, 1999}

\newsec{Introduction}

\def\bar{\overline}
Suppose that $M$ is a complete Einstein manifold of negative curvature
and dimension $n+1$
and that the conformal boundary of $M$, in the sense of Penrose
\ref\penrose{R. Penrose and W. Rindler, {\it Spinors And Spacetime},
vol. 2, chapter 9 (Cambridge University Press, Cambridge, 1986).},
is an $n$-manifold $N$.  This means that $M$ is the interior of
an $n+1$-dimensional manifold-with-boundary $\bar M$, whose boundary
is $N$, and that the metric $g$ of $M$ can be written near the boundary
as
\eqn\tugo{g={1\over t^2}\left(dt^2+g_{ij}(x,t)dx^idx^j\right),}
where $t$ is a smooth function with a first order zero on $\partial \bar M$,
and positive on $M$, and $g_{ij}(x,t)dx^idx^j$ is an $t$-dependent family of metrics on $M$.  Thus, $t\geq 0$ on $\bar M$ and $t=0$ on $N$.
In this situation,   $g_0=g(x,0)$ is a metric on $N$.
If $t$ is replaced by a different function with a first order zero
on $\partial\bar M$, say $t'=e^\omega t$, 
then $g_0$ undergoes a conformal transformation $g_0\to g_0'=e^{2\omega}g_0$,
so $N$ actually has a natural conformal structure but not a natural metric.
If, in the conformal class of metrics on $N$, there is a representative
with positive (or zero, or negative)
scalar curvature, then we say that $N$ has positive (or zero, or negative)
scalar
curvature.

There is a correspondence
\nref\maldacena{J. Maldacena, ``The Large $N$ Limit Of Superconformal
Field Theories And Supergravity,'' hep-th/9711200.}%
\nref\kleb{S. S. Gubser, I. R. Klebanov, and A. M. Polyakov,
``Gauge Theory Correlators From Non-Critical String Theory,''
hep-th/9802109.}%
\nref\witten{E. Witten, ``Anti de Sitter Space And Holography,''
hep-th/9802150.}%
between conformal field theory
on $N$ and quantum gravity, or string theory, on $M$ 
\refs{\maldacena - \witten}.
To be more precise, the correspondence asserts (see \witten, section 3)
that to do
conformal field theory on $N$ with a given conformal structure $g_0$ on $N$,
 one must sum over contributions of all
possible $n+1$-dimensional Einstein manifolds $M$ with conformal boundary
$N$ and induced conformal structure $g_0$.  Actually, the full
correspondence involves a number of additional details that we will
omit in the present paper.
For example, one usually must consider not $n+1$-dimensional Einstein
manifolds $M$, but manifolds of dimension $n+k+1$
obeying appropriate supergravity equations and 
asymptotic at infinity to $X= M\times Y$, where $Y$ is a fixed compact 
$k$-manifold
characteristic of the conformal field theory that one chooses to consider.
(Examples are given in \maldacena.)
Our results could possibly be extended to theorems about the possible
$X$'s (showing for example that under suitable hypotheses the 
ideal boundary of $X$ is connected),
 but for simplicity we will consider only the case that $X=M\times Y$,
and analyze the possible $M$'s.

\bigskip\noindent{\it Topological Conditions}

\def\AdS{{\rm AdS}}
Presented with this correspondence, one wonders how one can characterize
the $M$'s that exist for given $N$.  Are there general topological
conditions on $M$?  For some choices of $N$, can one find all of the $M$'s?
In this paper, we will prove the following general restriction:
if $N$ has positive scalar curvature (and is nonempty),
then $H_n(M;{\bf Z})=0$.  We explain
below the physical interpretation of this restriction.

The role of positive scalar curvature is  suggested by the
most canonical example of a complete, negatively curved Einstein manifold,
namely hyperbolic space ${\bf H}^{n+1}$, or -- as it is known in the 
supergravity literature  --
Anti de Sitter space $\AdS_{n+1}$.  In this case, the conformal
boundary $N$ is a sphere ${\bf S}^n$, with the standard conformal
structure which has as a representative the standard ``round'' metric.
This metric certainly has positive scalar curvature.  Because of the
basic role played by this example, the correspondence between quantum
gravity in $n+1$ dimensions
and conformal field theory in $n$ dimensions
is sometimes called the AdS/CFT correspondence.

\nref\malst{J. Maldacena, J. Michelson,  and A. Strominger, ``Anti de Sitter
Fragmentation,'' JHEP 9902:011,1999, hep-th/9804085.}
\nref\seiwit{N. Seiberg and E. Witten, ``The $D1/D5$ System And Singular
CFT,'' JHEP 9904:017,1999, hep-th/9903224.}
In general, conformal field theory can make sense on a manifold of
negative scalar curvature, but the specific conformal field theories
that arise in the AdS/CFT correspondence, at least in the examples
studied so far, are well-behaved only when the scalar curvature of $N$
is non-negative.  In the important case that $N$ is a four-manifold, 
this can be seen directly: in this case, the conformal
field theories are four-dimensional
gauge theories which contain scalar fields whose
potential is unstable if the scalar curvature of $N$ is negative.  
More generally, one sees by considering the action of 
a suitable brane in $M$ \refs{\malst,\seiwit} that if $N$ has negative
scalar curvature, the theory is unstable.  The argument, whose
details we recall in section 2, is made by considering a brane in
$M$ whose worldvolume is a codimension one hypersurface $\Sigma\subset M$.
One considers the brane action $L(\Sigma)$ (which we will define
in section 2) and shows that it
is unbounded below if $N$ has negative scalar curvature.  If $N$
has positive scalar curvature, the theory is stable, and if $N$ has
zero scalar curvature, it may be stable or unstable depending on further
details.

\bigskip\noindent{\it Two Puzzles Concerning The AdS/CFT Correspondence}

We therefore limit ourselves to the case that $N$ has positive
scalar curvature.  
Consider the following two puzzles
concerning the AdS/CFT correspondence:

(1) Can it happen that $N$ is not connected but is a union of
disjoint components $N_i,$ $i=1,\dots, s$, 
each of positive scalar curvature?
If so, the AdS/CFT correspondence does not make much sense.
For conformal field theory on a union of disjoint manifolds
$N=\cup_iN_i$ is just the product of the theories on the different
$N_i$'s.  There is no evident way to couple them, and hence no candidate
for how to interpret the contribution of an $M$ whose boundary
is $\cup_iN_i$. 

\nref\coleman{S. Coleman, ``Why There Is Nothing Rather Than Something:
A Theory Of The Cosmological Constant,'' Nucl. Phys. {\bf B310} (1988) 643.}%
\nref\banks{T. Banks, ``Prolegomena To A Theory Of Bifurcating Universes:
A Nonlocal Solution To The Cosmological Constant Problem, Or Little
Lambda Goes Back To The Future,'' Nucl. Phys. {\bf B309} (1988) 493.}%
(2) For given $N$ of positive scalar curvature, can $M$ contain 
``wormholes''?  A wormhole is obtained by cutting out two balls
from a manifold $M_0$ and then gluing together their boundaries
to make a new manifold $M$.  If $M$ can have such wormholes,
then to understand the conformal field theory on $N$, we will have
to come to grips with the strange behavior of quantum gravity in the
presence of wormholes \refs{\coleman, \banks}.

We will resolve both of these problems by showing that they simply
do not arise for $N$ of positive scalar curvature. 
  This follows from our result that for such $N$,
   $H_n(M;\Z)=0$.  Indeed, in (1), if the
number of boundary components is greater than one, then a hypersurface that is
near one of the boundary components (defined by an equation such as
$t=\epsilon$, where $t$ is the coordinate used in \tugo\ and
$t=0$ defines the boundary in question)
is a nonzero element of $H_n(M;\Z)$.  Hence vanishing of $H_n(M;\Z)$ implies
that $N$ is connected.  Likewise, in (2), 
a spacetime $M$ with wormholes would have nonzero $H_n(M;\Z)$, since
the boundary
of either of the balls removed from $M_0$ is a nonzero element of
$H_n(M;\Z)$.  Hence vanishing of $H_n(M;\Z)$ implies that there are
``no wormholes.'' 

Results of this type definitely depend on $N$ having positive scalar
curvature. For example, let $Q$ be any compact negatively curved
Einstein manifold of dimension $n$, with metric $g_{ij}dx^idx^j$.
Then a complete Einstein metric of negative curvature on $M=Q\times
{\bf R}$ is given by the simple formula
\eqn\simpform{ds^2= dt^2+\cosh^2t\,g_{ij}dx^idx^j.}
The conformal boundary of $M$ consists of two copies of $Q$, at $t=\pm \infty$.
This shows that if the conformal boundary of $M$
has negative scalar curvature, then $H_n(M;\Z)$ can be nonzero.  Note
that in this example, each component of $N$ has negative scalar curvature.
In fact, our proof shows that $H_n(M;\Z)=0$ if any component of $N$ has
positive scalar curvature.

An interesting corollary of the fact that, under the stated hypotheses,
$N$ must be connected is that the natural map from $\pi_1(N)$ to $\pi_1(M)$
is surjective.  (For example, $M$ must be simply connected if $N$ is
simply connected.)  Otherwise, by taking a suitable cover of $M$, one
could make an example with disconnected $N$ of positive scalar curvature.

\bigskip\noindent{\it Structure Of The Argument}

The proof that $H_n(M;\Z)=0$ for a boundary of positive scalar
curvature will be made by showing, by a local
calculation, that the brane action $L(\Sigma)$ cannot have a minimum,
and also by showing, using nonlinear analysis, that there is a minimum
in each nonzero homology class if the boundary has a component
of positive scalar curvature.
Combining these results, it follows that $H_n(M;\Z)=0$ if the boundary
has such a component.  The local computation is presented in section
2 and the global one in section 3.

\newsec{Local Calculation}

Let $M$ be an $n+1$-dimensional Einstein manifold of negative curvature
and nonempty conformal boundary, for some $n\geq 2$.
Let $\Sigma$ be a compact hypersurface in $M$.  We denote its area or
volume as $A$.  Since $M$ has a nonempty boundary, the volume form
$\Theta$ of $M$ is exact, say $\Theta=d\Lambda$ for an $n$-form
$\Lambda$.  The brane action (for a BPS brane) is defined by
\eqn\xono{L(\Sigma)= A - n\int_\Sigma \Lambda.}
Note that if $\Sigma$ is the boundary of a domain $\Omega$, we have
\eqn\bono{L(\Sigma) = A-n V}
where
\eqn\hono{V=\int_\Omega \Theta}
is the volume enclosed by $\Sigma$.  $\Lambda$ is not unique,
but changing $\Lambda$ will  add to $L$ a term that is
a constant in each homology class, and
this does not affect the variational problem that we will consider
below.  (In the physical application, $\Lambda$ is an $n$-form
field of the appropriate supergravity theory, and any choice of $\Lambda$
makes sense.)

The importance of positive curvature for the boundary is that it is
necessary in order to ensure that $L(\Sigma)$ is bounded below.
This follows from a computation performed in \malst\
and in greater generality in \seiwit. 
As in the introduction, we write the metric near the boundary
as
\eqn\metbound{ds^2={1\over t^2}\left(dt^2+g_{ij}(x,t)dx^idx^j\right).}
We consider $\Sigma$ to be homotopic to the boundary and to be defined
by an equation $t=t(x^i)$.  We
write
\eqn\udut{t=\cases{2\phi^{-2/(n-2)} & for $n>2$ \cr
                   2e^{-\phi} & for $n=2$.\cr}}
The brane action then becomes for small $t$ or equivalently large $\phi$
\eqn\hudut{L=\cases{{1\over 2^{n-3}(n-2)^2}\int_\Sigma d^nx \sqrt g\left(
|d\phi|^2+{n-2\over 4(n-1)}\phi^2R+{\cal O}(\phi^{2(n-4)/(n-2)})\right) &
 for $n>2$;\cr
 ~~~~
 {1\over 2}\int_\Sigma d^2x\sqrt g\left(|d\phi|^2+\phi R+{\cal O}(e^{-2\phi})
 \right) & for $n=2$.\cr}}
Here we have identified $\Sigma$ with the boundary at $t=0$, and
we regard $g_{ij}(x,0)$ as a metric $g_{ij}$ on $\Sigma$; $R$ is the scalar  
curvature of this metric.  
\foot{In deducing \hudut, one uses the Einstein equations to determine
the behavior of $g_{ij}(x,t)$ near $t=0$.  For details, see
\seiwit, eqns. (3.6)-(3.8).}
The brane action is conformally invariant;
indeed, under $t\to e^\omega t$ (where $\omega$ is a function on
the boundary, that is, on $\Sigma$), we have $g_{ij}\to e^{2\omega}g_{ij}$
and in view of \udut\
\eqn\nudut{\phi \to \cases{\exp(-(n-2)\omega/2)\phi& for $n>2$\cr
                           \phi-\omega   & for $n=2$.  \cr}}  
For $L$ to be bounded below, it must be bounded below in the region
of large $\phi$, where the corrections in \hudut\ can be dropped.
Whether this is so depends, for $n>2$, on the spectrum of the conformally
invariant operator 
\eqn\dobo{\bigtriangleup'=\bigtriangleup +{n-2\over 4(n-1)}R,}
where $\bigtriangleup$ is the Laplacian.  If $\bigtriangleup'$ is
positive definite, $L$ is bounded below at least in the region of large
$\phi$; if it has a negative eigenvalue, then $L$ is unbounded below;
and if the smallest eigenvalue of $\bigtriangleup'$ is zero, then
one must consider the correction terms in \hudut\ to determine if
$L$ is bounded below in the large $\phi$ region.  
The lowest eigenvalue of $\bigtriangleup'$ is positive,
negative, or zero depending on whether, in the conformal class of
the metric $g_{ij}$ on $\Sigma$, there is a representative of
positive, negative, or zero scalar curvature. 

Thus we learn that, for $n>2$, stability requires that $\Sigma$ have
nonnegative scalar curvature.  For $n=2$, the same conclusion can be
reached by first replacing $g$ with a conformally equivalent metric
$e^{2\omega}g$ with constant $R$, and then noting that for constant
$\phi$ and $\phi\to +\infty$, $L$ is bounded below if $R\geq 0$ and
unbounded below if $R<0$.  

\bigskip\noindent{\it Properties Of A Minimum Of $L$}

So far we have merely summarized the considerations in
\refs{\malst,\seiwit}.
Now, assuming that the boundary of $M$ has positive scalar
curvature, we want to try to get a restriction on the topology of $M$.
The computation performed above suggests that
(for a boundary of positive scalar curvature)
 $L$ is bounded below.  If so, we may expect that
$L$ will have an absolute minimum for each nonzero choice of the
 homology class of $\Sigma $ in $H_n(M;{\bf Z})$.  (The reason
that one suspects a minimum for each homology class, not each homotopy
class, is that in varying a hypersurface $\Sigma$ to minimize $L$,
$\Sigma$ may develop a singularity.  In passing through such a singularity,
the homology class of $\Sigma$ does not change, but the homotopy class
may.)  The existence of such a minimum will be proved in section 3.
We will now show, however, by a local computation that $L$ cannot have
a minimum.  Combining these results, it will follow that
$H_n(M;{\bf Z})=0$, which is the result that was promised in the introduction.

We assume that the Einstein manifold $M$ obeys
\eqn\imco{R_{IJ}=-ng_{IJ}.}
(The choice of constant on the right hand side
 is correlated with the choice of constant
$n$ multiplying the second term in the brane action \bono.
We denote indices of $M$ by $I,J,K=1,\dots,n+1$ and indices of the
hypersurface $\Sigma$ by $i,j,k=1,\dots,n$.)

\def\hat{\widehat}
We now want to study a hypothetical extremum $\Sigma$ of $L$.  We suppose
that $\Sigma$ is an embedded submanifold.  
In a neighborhood of $\Sigma$, we can pick one
coordinate, $r$, to be the distance from $\Sigma$, and pick the other 
coordinates
$x^i$ so that the metric near $\Sigma$ is
\eqn\loopy{ds^2=dr^2+\hat g_{ij}(x,r)dx^idx^j.}
We let $g_{ij}=\hat g_{ij}(x,0)$ be the metric on $\Sigma$, and we write
\eqn\oopy{\eqalign{ \dot g_{ij}=&\left.
{\partial \hat g_{ij}(x,r)\over \partial r}\right|_{r=0} \cr
     \ddot g_{ij}= & \left.{\partial^2\hat g_{ij}(x,r)
\over \partial r^2}\right|_{r=0}.\cr}}
(Thus, the second fundamental form of $\Sigma$ is $\dot g/2$, and
the mean curvature is $\Tr\,g^{-1}\dot g/2$.)

  We can describe a fluctuation in
the position of $\Sigma$ by specifying $r$ as a function of $x$.  
To determine the conditions
for $\Sigma$ to be a local minimum of $L$, we need to evaluate $L$ up to
second order in $r$.  
Writing $\widehat g=g+r\dot g+\half r^2\ddot g+\dots $, 
we calculate to this order
\eqn\huvn{\eqalign{ A=
\int d^nx\sqrt{\det (\hat g_{ij}+\partial_ir\partial_jr)}
=&\int d^nx\sqrt{\det g}\left(1+{r\over 2}\Tr\,g^{-1}\dot g+{r^2\over 8}
(\Tr\,g^{-1}\dot g)^2\right.\cr &\left.
+{r^2\over 4} \Tr\,g^{-1}\ddot g-{r^2\over 4}
\Tr(g^{-1}\dot gg^{-1}\dot g)+ \half |dr|^2\right).\cr}}
Also, to this order
\eqn\buvn{nV={\rm const}+n\int d^nx
\sqrt g\left(r+{r^2\over 4}\Tr\,g^{-1}\dot g
\right).}
So
\eqn\puvn{\eqalign{L=A-nV={\rm const}+\int d^nx\sqrt g&\left({r\over 2}
\Tr\,g^{-1}\dot g-rn
+{r^2\over 8}(\Tr\,g^{-1}\dot g)^2-{nr^2\over 4}\Tr\,g^{-1}\dot g \right.\cr
&\left.
+{r^2\over 4}\left(\Tr\,g^{-1}\ddot g-\Tr(g^{-1}\dot gg^{-1}\dot g)\right)
+\half |d r|^2\right).\cr}}
The condition for $L$ to be stationary at $r=0$ is
\eqn\ucbu{\Tr \,g^{-1}\dot g=2n.}
For $r=0$ to be a local minimum  requires
\eqn\bucbu{\Tr\,g^{-1}\ddot g\geq \Tr(g^{-1}\dot gg^{-1}\dot g).}
If $N$ is a real symmetric $n\times n$ matrix,
then
\eqn\matform{\Tr\,N^2\geq {1\over n}(\Tr\,N)^2,} with equality
only if and only if $N$ is a multiple of the identity.
Applying this to $N=g^{-1}\dot g$, for which  $\Tr \, N=2n$, 
we learn that
\eqn\jucubu{\Tr\,(g^{-1}\dot gg^{-1}\dot g)\geq 4n,}
with equality precisely if $\dot g = 2g$.

Now we look at the Einstein equations.  If $\Gamma$ are the ($r$-dependent)
Christoffel
symbols of $\Sigma$ in the  metric $g_{ij}(x,r)dx^idx^j$, then the nonzero
Christoffel symbols $\hat \Gamma$ of $M$ in the metric \loopy\ are
\eqn\nubbob{\eqalign{\hat\Gamma^i_{jk}& = \Gamma^i_{jk}\cr
                         \hat \Gamma^r_{jk}& = -\half \dot g_{jk}\cr
                           \hat\Gamma^i_{rj} & =\half g^{is}\dot g_{js}.\cr}}
Let $R_{ij}$ and $R$ be the Ricci tensor and scalar of $\Sigma$, and
$\widehat R_{ij}$, $\widehat R$ the analogous objects of $M$.
The relevant part is
\eqn\eqnubo{  \hat R_{rr}  = -{1\over 2} \Tr\, g^{-1}\ddot g +{1\over 4}
  \Tr\, g^{-1}\dot gg^{-1} \dot g        }
Now we use the Einstein equations at $r=0$;
the  equation $\hat R_{rr}=-ng_{rr}=-n$ gives
\eqn\nucovon{\half\Tr\,(g^{-1}\ddot g)-{1\over 4}\Tr \,(g^{-1}\dot gg^{-1}
\dot g)= n.}    
Using also the inequality \bucbu\ that followed from stability,
we get 
\eqn\mocov{4n\geq \Tr\,(g^{-1}\dot gg^{-1}\dot g).}  
Comparing to \jucubu, we learn that all the inequalities must be equalities,
forcing $\dot g=2g$, $R=0$, and $\Tr\,g^{-1}\ddot g=4n$.

So the possibility that the action $L$ has a stable nondegenerate
minimum is excluded.   Note that
the analysis has been purely local and makes
no assumption about the global structure of $M$.
 
As for the case of a degenerate minimum where
the order $r^2$ term vanishes, a further analysis that we will
explain momentarily
shows that this can happen only in an example of the 
following type. If the metric $g_{ij}$ on $\Sigma$ is
Ricci-flat, then the metric
\eqn\jxcon{ds^2=dr^2+e^{2r}g_{ij}dx^idx^j}
on $\R\times \Sigma$
obeys the $n+1$-dimensional Einstein equations with cosmological constant.
For any constant $c$,
the submanifold $\Sigma_c$ of $\R\times \Sigma$
  defined by $r=c$ is a stationary point
of $L$.  The  action  $L(\Sigma_c)$ is independent of $c$, so this
is a degenerate critical point.  Conformal infinity
consists of a copy of $\Sigma$ at $r=\infty$, with zero scalar curvature.
Thus this type of example is impossible if we assume that the boundary
has positive scalar curvature.
(In this type of example, 
there is also a sort of cusp at $r=-\infty$, so there is no
Penrose compactification even with nonpositive curvature on the boundary.)

We conclude by giving the proof that a degenerate minimum of $L$
must be of the form just described.  Let  $L(c)=L(\Sigma_c)$.
We have from \puvn\
\eqn\ixcono{{dL\over dc}= -\int_{\Sigma_c}d^nx \sqrt g F}
where
\eqn\bixco{F=n-{1\over 2}\Tr\,\, g^{-1}\dot g.}
Since we assume that $L$ is locally minimized at $c=0$, we have
$dL/dc\geq 0$ for small positive $c$, and hence
\eqn\mixco{\int d^nx \sqrt g F \leq 0.}
On the other hand,
\eqn\jixco{{d\over dr}F={1\over 2}\Tr(g^{-1}\dot gg^{-1}\dot g)
-{1\over 2}\Tr\,\, g^{-1}\ddot g.}
Using \nucovon, this becomes 
\eqn\hixco{{dF\over dr}=-{n}+{1\over 4}\Tr(g^{-1}\dot gg^{-1}\dot g).}
Using \matform, with $N=g^{-1}\dot g$, 
this implies an inequality
\eqn\uju{{dF\over dr}\geq -{1\over n}F(2n-F).}
We have $F(0)=0$, since $\Sigma_0$ is a critical point of $L(\Sigma)$.
For small positive $r$, $F$ has the same sign as $F'$  
since $F(r)=\int_0^rdt\, F'(t).$   \uju\ implies that if $F(r)$ is negative
for small positive $r$, then  $dF/dr$ is positive.
This is a contradiction, so $F\geq 0$ for small positive $r$.

Comparing to \mixco, we learn that $F$ is identically zero 
for all sufficiently
small positive $r$.  It follows from \ixcono\ that $L(c)$ is independent
of $c$ for small positive $c$.  As we deduced from \mocov, at any
value of $c$ for which $dL/dc=d^2L/dc^2=0$, we have
$dg_{ij}/dc=2g_{ij}$.  Hence 
\eqn\tufco{g_{ij}(x,c)=e^{2c}g_{ij}(x,0)}
for
sufficiently small positive $c$.   By real analyticity (or the
Einstein equations) this is true for all $c$, and the Einstein
equations \imco\ also imply that $g_{ij}(x,0)$ is a Ricci-flat metric
on $\Sigma$.  Thus, we have shown that a degenerate minimum of the
functional $L(\Sigma)$ has the special form given in \jxcon\
and in particular cannot exist if the boundary has positive
scalar curvature.

In the above, we can replace the Einstein equation $R_{IJ}=-ng_{IJ}$
by an inequality $R_{IJ}\geq -ng_{IJ}$, since this would only
improve the crucial inequality \nucovon.
Physically, this corresponds to having additional matter fields
excited in an asymptotically AdS spacetime.

We summarize our results as follows:

\noindent{\bf Theorem 2.1} {\it The functional $L(\Sigma)=A-n\int_\Sigma
\Lambda$
for an embedded hypersurface $\Sigma$ in an $n+1$-dimensional
Einstein manifold $M$ of Ricci curvature greater than or equal to
$-n$ does not have any
local minimum.  Any critical point of this functional is either
unstable or is neutrally stable and of the form given in \jxcon.  The
neutrally stable case is only possible if the Ricci curvature is
precisely $-n$.}

\newsec{Existence}

In this section, we will prove existence theorems for
a hypersurface $\Sigma$, in a given homology class,
that minimizes the functional $L(\Sigma)={\rm Area}(\Sigma)
-n\int_\Sigma\Lambda$.  Here $\Sigma$ is a hypersurface in
a complete $n+1$-dimensional manifold $M$ of Ricci curvature $-n$
that has a conformal
boundary as described in the introduction and the last section.

\nref\gonzalez{E. Gonzalez, U. Massari, and I. Tamanimi,
``On The Regularity Of Boundaries Of Sets Minimizing Perimeter
With A Volume Constraint,'' Indiana U. Math. J. {\bf 32} (1983) 25.}
\nref\gru{M. Gr\"uter, {\it Boundary Regularity For Solutions Of
A Partitioning Problem}, Arch. Rat. Mech. Anal. {\bf 97} (1987) 261.}
\nref\simon{L. Simon, {\it Lectures On Geometric Measure Theory},
Proc. Center for Math. Anal., Australian Nat. Univ. {\bf 3} (1983).}
\nref\sz{P. Sternberg and K. Zumbrun,
``On The Connectivity Of Boundaries Of Sets Minimizing Perimeter
Subject To A Volume Constraint,'' Comm. Anal. Geom. {\bf 7}
(1999) 199.}

If instead $M$ were compact, and we have an upper bound on the
area of $\Sigma$ and a lower bound on $L(\Sigma)$, then existence
of a minimizing hypersurface 
$\Sigma$ in a given homology class in $M$
follows from very general grounds.  In fact, any sequence of
hypersurfaces of bounded area in a compact manifold has a convergent
subsequence.  For compact $M$, if we assume that $\Sigma$ is
an embedded hypersurface, then both the upper bound on the area
in a minimum of $L(\Sigma)$ and the lower bound on $L(\Sigma)$ follow
from the fact that the potentially negative term $-n\int_\Sigma \Lambda$
in the definition of $L(\Sigma)$ is bounded below by minus
the volume of $M$.
So if we make  $M$ compact by cutting off the ``ends,'' then $\Sigma$ exists.

The $\Sigma$ obtained this way, as the limit of a sequence of embedded
hypersurfaces $\Sigma_i$ chosen to minimize $\lim_{i\to\infty}L(\Sigma_i)$,
might {\it a priori} have very bad singularities.  However,
rather deep results in geometric measure theory
\refs{\gonzalez - \sz}  show that such a limiting $\Sigma$
has singularities only in codimension $\geq 7$.
(These results are obtained for area-minimizing hypersurfaces.
The possible singular behavior at interior points of $M$
of a hypersurface minimizing
$L=A-n\int \Lambda$ 
 is the same as for area-minimizing hypersurfaces, 
since the second term is less important
near a singularity.)  Existence of a minimizing
hypersurface $\Sigma$ that is smooth except in high
codimension is good enough for our purposes, because the arguments
of section 2, though formulated for smooth $\Sigma$, can be extended
to the case that $\Sigma$ has a singularity of high codimension.
\foot{Curiously, in supersymmetric examples of the
AdS/CFT corresponce (and in fact, in all known examples) the
dimension of $M$ is at most seven and hence the dimension of $\Sigma$
is at most six.  So in the known applications, $\Sigma$ is always smooth.}

To apply this existence result for $\Sigma$
to the case that $M$ is noncompact and has
a Penrose compactification,
we first introduce a cutoff in the volume of $M$
as follows.  We recall that near each conformal boundary
component $\partial M_i$ of $M$, the metric of
$\Sigma$ looks like
$$ {1\over t^2}\left(dt^2+g_{ij}(x,t)dx^idx^j\right), $$
with the boundary being at $t=0$.  We cut off the ``ends'' by restricting
to $t\geq \epsilon_i$, with $\epsilon_i$ a small positive function on $\partial
M_i$.  With $M$ made compact in this way, existence of $\Sigma$ follows
by the argument above.  The main technical step
in the present section is to prove, under certain conditions,
that if the $\epsilon_i$ are sufficiently small, then the minimizer
$\Sigma$ (or at least one of its components) does not intersect
the boundary of $M$.  Once this is known, a comparison with
Theorem 2.1 will give our
restrictions on the topology of $M$.

\def\al{\alpha}
\def\Om{\Omega}
\def\Si{\Sigma}
\def\si{\sigma}
\def\ve{\varepsilon}

\def\pa{\partial}
\def\ti{\times}
\def\ra{\rightarrow}
\def\wt{\widetilde}
\def\s{\sqrt}

We turn now to the proofs. 
We begin with some preparations. The following is a well-known fact:

\noindent
{\bf Lemma 1.}
{\it Let $d$ be the geodesic distance function from a point $x_0$ in a
manifold $M$ whose sectional curvature has an upper bound given by
$k>0$. At points where $\sqrt{k}d<\pi/2$ and $d$ is smooth, the
second derivative (the Hessian) of $d$, in directions orthogonal to
the tangent vector of the shortest geodesic joined to $x_0$, is not
less than
\eqn\twoptone{
{1\over d}{{\sqrt{k}d}\over {\rm tan}\sqrt{k}d}
.}}

\noindent
{\it Proof.} Let $\si(s,t):(-a,a)\ti[0,l]\ra M$ be a family of geodesics so
that
$$
\si(s,0)=x_0\qquad {\rm for~ all }~ \;s,
$$
and at $s=0$, ${d\over ds}\si(s,l)$ is perpendicular to ${d\sigma\over dt}(0,l)$. We also suppose that $\si(s,l)$ is a geodesic.
We shall parametrize $\si(0,t)$ by arc length so that 
$t={\rm length}\,\si(0,t)$. Then $J(t)={d\over ds}\si(s,t)\Big|_{s=0}$
is perpendicular to ${d\si\over d t}$ for all $t$. Direct
calculation shows that
\eqn\twopttwo{{d^2\over ds^2}\,{\rm length}\,\si(s,t)\Big|_{s=0}
=\int_0^l\left[\left|{d\over dt}J\right|^2
-\int_0^lR_{ijkl}J^iJ^k{{d\si^j}\over{dt}}\,{{d\si^l}\over{dt}}\right]dt.}

Assume the sectional curvature $\Si R_{ijkl}X^iX^kY^jY^l$ to be less
than $k\|X\|^2\|Y\|^2$ when $X\perp Y.$ Then we see that
\eqn\twoptthree{
{d^2\over ds^2}\,{\rm length}\,(\si)\Big|_{s=0}
\geq\int_0^l\left\|{{dJ}\over{dt}}\right\|^2
-k\int_0^l\|J\|^2.}

Assuming $\sqrt{k}l\leq{{\pi}\over {2}}$, the right hand side is minimized
by setting 
\eqn\twoptfour{
J(t)={{\sin(\sqrt{k}t)}\over{\sin(\sqrt{k}l)}}J(1).}

Hence 
\eqn\twoptfive{
{d^2\over ds^2}\,{\rm length}\,(\si)\Big|_{s=0}
\geq {{\sqrt{k}}\over {\tan(\sqrt{k}l)}}.}

Let $\rho=d^2$. Then the second derivative of $\rho$ is not less than
\eqn\twoptsix{
{{2\sqrt{k}d}\over{\tan(\sqrt{k}d})}}
if the derivatives are taken along directions orthogonal to the
shortest geodesic joint to $x_0$. It is not less than 2 if the
direction is tangential to the shortest geodesic joint to $x_0$. Hence
the second derivative is always not less than 
\eqn\threeseven{
f(\rho)=2\min\left({{\sqrt{k}d}\over
{\tan(\sqrt{k}d)}},\;1\right).}

Let $\Si$ be a hypersurface with mean curvature bounded above
by $c$.\foot{In the notation of section 2, the mean
curvature is $\Tr\, g^{-1}\dot g/2$.}
 Then 
restricting $\rho$ to $\Si$, we find
\eqn\threeeight{
\Delta_{\Si}\rho\geq nf(\rho)-2c\sqrt{\rho}.}

This follows because the second derivatives on $\Si$ differ from the
ones on $\Si$ by the second fundamental form after being renormalized by
the gradient of $\rho$, whose norm is less than $2d=2\sqrt{\rho}$.
(The first derivative of the geodesic distance is not greater than one.)

Let us now assume that distance from $x\in\Sigma$ to $\pa M$ is greater
than $\sqrt{R}$. 
Integrating the above inequality on $\Si\cap\{\rho\leq R\}$, we
obtain
\eqn\threenine{
2\sqrt{R}\,{\rm Area}\,\left[\Si\cap\{\rho=R\}\right]
\geq\int(nf(\rho)-2c\sqrt{\rho}).
}
(Here we have used the fact that, since the
normal derivative of $d$ is at most 1, 
the normal derivative of $\rho$ is not greater than $2\sqrt{R}$ when 
$\rho=R$.)

Let $F(R)={\rm Volume}\,\,[\Si\cap\{\rho\leq R\}].$ Then
\eqn\twoten{\eqalign{
{{dF}\over {dR}} 
&\geq {{1}\over {2\s{R}}}\,{\rm Area}\,\left[\Si\cap\{\rho=R\}\right]
\cr
&\geq\int_{\rho\leq R}\left({{nf(t)-2c\s{t}}\over {4R}}\right)\cr
&\geq\inf_{t\leq R}\left({{nf(t)-2c\s{t}}\over {4R}}\right)F(R).
\cr}}

In particular,
\eqn\threeeleven{
(\log F(R))'\geq (\log H(R))'
}
where
\eqn\threetwelve{
H(R)=R^{n/2}\exp\left\{\int_0^R\left(\inf_{s\leq t}
\left({{nf(s)-2c\s{s}}\over {4t}}\right)-{{n}\over
{2t}}\right)\right\}.}
When $R\ra 0,$ $H(R)\ra R^{n/2}$ and $F(R)\sim CR^{n/2}.$ Hence
\eqn\threethirteen{
F(R)\geq CR^{n/2}\exp\left\{\int_0^R\left(\inf_{t\leq R}
\left({{nf(t)-2c\s{t}}\over{4R}}\right)-{{n}\over {2t}}\right)\right\}.
}

\noindent
{\baselineskip=0pt
{\bf Lemma 2.}
{\it Let $\Si$ be a hypersurface with a mean curvature bounded above 
by $c$ in a manifold $M$ whose sectional 
curvature is bounded above by a constant $k$. Let 
$x_0\in\Si$ be a point so that the distance from $x_0$ to $\pa M$ is
greater than $R$. Then if $R<{{\pi}\over {2\s{k}}}$ and if the geodesic 
distance (of $M$) from $x_0$ is smooth within $B_{x_0}(R)$ the ball of
radius $R$, then the area of $B_{x_0}(R)\cap\Si$ is not less than 
$\wt{c}R^n$ where $\wt{c}$ depends only on $\dim M$, $\s{k}R$ and
$cR$.}}

The assumption that the geodesic function from $x_0$ is smooth will be
true if $R$ is small enough. This can be seen as follows. The only
reason that the geodesic distance
 may  not be smooth at $x$ is that there may be two distinct
geodesics with shortest distance joining $x$ to $x_0$. By minimizing the
distance of such geodesics, one can find a point $x$ so that the
distinct geodesics at $x$ have exactly opposite directions and hence
there is a smooth geodesic loop at $x_0$.

Let us now assume that for some constant $\al>0$, the map
$\pi_1(B_{x_0}(\al R))\ra \pi_1(B_{x_0}(R))$ is trivial. Since
$R<{{\pi}\over {2\s{k}}}$, the exponential map from the tangent space at
$x_0$ is non-singular everywhere in a ball of radius $R$.

The geodesic loop at $x_0$ bounds a disk within $B_{x_0}(R)$ and hence
can all be lifted up to the tangent space.  This is not possible as the
lifting of any geodesic at $x_0$ must be a straight line and cannot be
a closed loop.

The condition for smoothness of geodesic distance is therefore
satisfied if we can find $\al>0$ so that $\pi_1(B_{x_0}(\al R))\ra
\pi_1(B_{x_0}(R))$ is trivial for all $0<R<{{\pi}\over {2\s{k}}}$ and for
all $x_0$ with distance greater than $R$ from $\pa M$. This condition
is clear for manifolds with compactification of the type described in
section 1.

{}From now on, we shall choose $R$ so that all the above assumptions are
satisfied. 

Suppose $\{x_1,x_2, \ldots, x_m\}$ are points on $\Si$ so that the balls
$B_{x_i}({{R}\over {2}})$ are mutually disjoint and $\Si\subset\cap_i
B_{x_i}(R)$. We also assume distances from $x_i$ to $\pa M$ is greater
than ${{R}\over {2}}$.  Then ${\rm Area}\,(\Si)\geq\sum_i{\rm Area}\,
(B_{x_i}({{R}\over {2}})
\cap\Si)$. Since each Area$\,(B_{x_i}({{R}\over {2}})\cap\Si)$ 
is bounded from below by a positive constant depending only on $R$,
$c$ and $\s{k}$, we conclude that $m$ is dominated by Area$\,(\Si)$. 
This number $m$ can be considered as a quantity that measures the 
outer diameter of $\Si$.

Let us now assume that $M$ is a compact manifold with boundary
components $\pa M_1,\ldots,\pa M_k.$ We assume $k>1.$ We
consider domains $\Om_{\Si}$ with boundary components given by 
$\pa M_2,\ldots,\pa M_k$ and an embedded hypersurface $\Si$ which is
(compactly) homologous to $\pa M_1.$ Then
\eqn\twofourteen{\eqalign{
L(\Si) &={\rm Area}\,(\Si)-n\int_{\Si}\Lambda\cr
&={\rm Area}\,(\Si)+\sum_in\int_{\pa M_i}\Lambda
-n\,{\rm Vol}\,(\Om_{\Si}).\cr}}

If $\Si_0$ minimizes $L(\Si)$,
\eqn\threefifteen{{\rm Area}\,(\Si_0)-n\,{\rm Vol}\,(\Om_{\Si_0})
\leq {\rm Area}\,(\pa M_1)-n\,{\rm Vol}\,(M).
}

Assume that $\Si_0$ can be written as $\Si_1+\cdots+\Si_k$. Then
either one of $\Si_i$ is in the interior of $M$ or all of them
intersect $\pa M.$ In case all of them intersect $\pa M$, we argue
as follows.  We consider first the case that all of the $\Si_i$ intersect
one of the $\partial M_j$ with $j>1$.
 From the above inequality (3.15), we know 
\eqn\threesixteen{\sum_i{\rm Area}\,(\Si_i)\leq {\rm Area}\,(\pa M_1).
}

By Lemma 2, the diameter of each $\Si_i$ is dominated by 
Area$\,(\Si_i)$ and hence by 
$a\,{\rm Area}\,(\pa M_1)$ where $a$ depends only on the upper bound
of the sectional curvature of $M$ and the lower bound of the injectivity
radius of $M$.   (The mean curvature of each $\Si_i$ is $n$, according
to \ucbu.) 

If $\Om$ is the complement (in $M$) of the neighborhood of $\cup_{j>1}(\pa
M_j)$ with radius of $a\,{\rm Area}\,(\pa M_1)$, then (as we are assuming
that each $\Sigma_i$ intersects one of the $\partial M_j$ with $j>1$) 
$\Om\subset(M\backslash\Om_{\Si_0})$ and (3.15) shows
\eqn\twoseventeen{
\sum_{i\geq 1}{\rm Area}\,(\Si_i)\leq {\rm Area}\,(\pa M_1)
-n\,{\rm Vol}\,(\Om).}
 In particular, Area$\,(\pa M_1)
+ n {\rm Vol}\,(M\backslash\Om) \ge n\,{\rm Vol}\,(M).$

We have then proved the following lemma:

\noindent{\bf Lemma 3.}
{\it On a compact manifold $M$ 
with boundary $\pa M_1,$  
$\pa M_2,\ldots,\pa
M_k$, let $\Si_1+\cdots+\Si_i$ be a sum of embedded cycles that
minimizes the functional $L(\Si)={\rm Area}\,(\Si)-n\,{\rm Vol}\,
\Om(\Si,\pa M_2,\ldots, \pa M_k)$, on the homology class of $\pa
M_1$. Then one of the $\Si_i$'s does not intersect $\pa
M_2\cup\cdots\cup\pa M_k$ if 
\eqn\threeeighteen{n\,{\rm Vol}\,M>{\rm Area}\,(\pa M_1)
+n\,{\rm Vol}\,B_d(\pa M_2\cup\cdots\cup\pa M_k)}
where $d=a\,{\rm Area}\,(\pa M_1)$, $B_d$ is the ball of radius $d$
around $\pa M_2\cup\cdots\cup\pa M_k$ and $a$ depends only on the
upper bound of the sectional curvature of $M$ and the lower bound of
its injectivity radius.
}

In the case that $M$ is obtained from a manifold with a Penrose
compactification by ``cutting off'' the ends by $t\geq \epsilon_i$
for $i\geq 2$, the inequality in Lemma 3 is obeyed if the $\epsilon_i$
are sufficiently small.

We still need a condition to ensure that a component of $\Sigma$ does
not meet $\pa M_1$.  When $\pa M_1$ is a conformal boundary at infinity,
the computation at the beginning of section 2 suggests that $\Sigma$
will not be near $\pa M_1$ if
 $\pa M_1$ has positive scalar curvature; we want to make
this more precise.  In the present
discussion, we have cut off the ends of $M$ and $\pa M_1$ is an ordinary
boundary; positive scalar curvature at conformal infinity implies
that (when $\pa M_1$ is sufficiently close to infinity) the mean
curvature of $\pa M_1$ is greater than $n$.

Thus in general the assumption we want is
that $\pa M_1$ has mean curvature greater
than $n$.
There is a foliation in a neighborhood of $\pa M_1$ so that the leaves 
are given by level sets of the distance function to $\pa M_1$. The
mean curvature of the nearby level sets is still greater 
than $n.$ The (outer) normal of these level sets defines a vector field $v$ 
in a neghborhood of $\pa M_1$ whose divergence is given by the mean 
curvature of the level sets. Hence
\eqn\threenineteen{{\rm div}\,v>n.}

Now if $\Si_i$ intersects $\pa M_1$, we can replace $\Si_i$ by intersecting 
it with the level sets and obtain a new surface $\widetilde{\Si_i}$.
By applying the divergence theorem (as norm$(v)=1$)
to the domain $\Om$ bounded by the difference of $\Si_i$ and 
$\widetilde{\Si_i}$, we get
\eqn\twotwenty{\eqalign{
{\rm Area}\,\widetilde{\Si_i}
&\leq {\rm Area}\,\Si_i-\int_{\Om}{\rm div} \,v\cr
&<{\rm Area}\,\Si_i-n\,{\rm Vol}\,(\Om).\cr}}

{}From this inequality, it is clear that $L(\widetilde{\Si_i})<L(\Si_i).$
Hence $\Si_i$ cannot be part of the minimum of the functional $L$.
In conclusion, if $\pa M_1$ has mean curvature strictly greater than $n$, 
it does not intersect any of the $\Sigma_i$, and hence under the
hypotheses of Lemma 3,
there must be a component $\Si_i$ which does not intersect any boundary 
components $\pa M_j$.  We can deduce the following:

\noindent{\bf Lemma 4.} {\it 
Let $M$ be a 
compact manifold with boundary components $\pa M_1,\ldots, \pa M_k$.
Assume that $\pa M_1$ has mean curvature greater than $n$. Let 
$B_d(\pa M_2,\ldots, \pa M_k)$ be a neighborhood of $\pa M_2\cup\cdots
\cup\pa M_k$ with radius $d=a\,{\rm Area}\,(\pa M_1)$, where $a$ depends
only on the upper bound of sectional curvature of $M$ and the lower bound 
of the injectivity radius of $M$. Assume that
\eqn\twopttwentyone{n\,{\rm Vol}\,[M\setminus B_d(\pa M_2,\ldots,\pa M_k)]
> {\rm Area}(\pa M_1).
}

Then when we minimize the functional $L(\Si)={\rm Area}\,(\Si)-
n\int_{\Si}\Lambda$ among embedded surfaces homologous to $\Si$
which bound a domain with $\pa M_2\cup\cdots\cup\pa M_k$, there must be
a component of $\Si$ which does not touch $\pa M_1\cup\cdots\cup\pa M_k$.
The singular set of $\Si$ is a closed set with at least codimension seven
Hausdorff dimension.}

This last statement follows from regularity theorems of geometric measure
theory. The fact that the singular set has large Hausdorff codimension allows
us to use arguments of section two. What one needs is to introduce
in the computations in section 2 a cut off function
$\varphi$ which is zero on the singular set and one outside an 
$\ve$-neighborhood of the set. The contribution of $\nabla\varphi$
is ${{1}\over {\ve}}$. But the singular set has small measure and the integral 
of $|\nabla\varphi|^2$ tends to zero when $\ve\ra 0.$

If the number of $\partial M_j$ is greater than one, so that $\partial M_1$
is not homologous to zero, then a minimum of $L(\Sigma)$ exists
in the homology class of $\partial M_1$ (as explained at the beginning
of this section) and Lemma 4 gives a condition in which the minimum
has a component that does not intersect the boundary of $\Sigma$.
But the existence of such a component contradicts Theorem 2.1 if the
Ricci curvature of $M$ is not less than $-n$.  So we conclude:

\noindent
{\bf Theorem 3.1} {\it
Let $M^{n+1}$ be a compact manifold with Ricci curvature not less  than
$-n$.  Let $\pa M_1$ be one of the boundary components of $M$ so that
$R_{\pa M_1} - R_M > {{1}\over {2}}n(n+1)$ along $\pa M_1$.  Assume that
\eqn\threetwentytwo{n {\rm Vol} [M \setminus B_d(\pa M_2, \ldots, \pa M_k)]
> {\rm Area}(\pa M_1),}
where $d$ depends only on the upper bound of the sectional curvature of $M$
 and
the lower bound of the injectivity radius. [$B_d(\pa M_2, \ldots, \pa M_k)$
is the neighborhood of radius $d$ around the components of $\pa M \setminus
\pa M_1$.] Then $\pa M$ has only one boundary component.}

 The above inequality on the volume is true if the boundary components
$\pa M_i$, $i \ge 2$, are far away from $\pa M_1$.
If $M$ is obtained by cutting off the ``ends'' in a Penrose compactification,
we can obey this inequality by moving the $\pa M_i$, $i\geq 2$, close enough
to infinity.

In the statement of Theorem 3.1, the condition 
$R_{\pa M_1} - R_M > {{1}\over {2}}n(n+1)$ ensures that the mean
curvature of the boundary is greater than $n$.  

Hence we have:

\noindent
{\bf Theorem 3.2.} {\it
Let $M^{n+1}$ be a complete manifold without boundary
with Ricci curvature not less than $-n$ and with a Penrose
compactification such that at least one component of the conformal
boundary of $M^{n+1}$ has positive scalar curvature.
Then the conformal boundary of $M^{n+1}$ is connected. 
More generally, let $M^{n+1}$ be any complete manifold
with curvature  bounded from above
and with Ricci curvature not less
than $-n$. Suppose $\pa M$ is
compact and $R_{\pa M}-R_M> {{1}\over {2}}n(n+1)$ along $\pa M$. If the
injectivity radius of $M$ is bounded from below by a positive constant,
then $M$ is compact, with connected boundary $\pa M$.
}

\bigskip

\noindent{\it Proof.}
If $M$ is not compact, we can exhaust $M$ by subdomains whose boundary
is $\pa M$ and $\pa M_2,\ldots,\pa M_k.$ We can make $(M\,\backslash$
$B_d(\pa M_2,\ldots,\pa M_k))$ to be arbitrarily large because $M$ is
complete and noncompact and we can put an arbitrarily large number of balls
$B_{x_i}(R)$ in $M\,\backslash\,B_d(\pa M_2,\ldots,\pa M_k)$ as long as
the subdomain is large.  Hence, we can obey the inequality in Lemma 4.

\noindent
{\bf Theorem 3.3} {\it
Let $M$ be as in Theorem 3.2.  Then the natural map
$\pi_1(\pa M)\ra \pi_1(M)$ is
surjective.
}

\noindent{\it Proof.}
Otherwise a non-trivial element of $\pi_1(M)\backslash \pi_1(\pa
M)$ exists. We can form a covering manifold $\wt{M}$ of $M$ making use
of this element. This $\wt{M}$ would have disconnected components.

\noindent
{\bf Theorem 3.4} {\it
Let $M$ be as in Theorem 3.2. Then $H_n(M;\Z)$ is zero.
}

\noindent{\it Proof.}
Let $\Si_0$ be a fixed embedded cycle representing an element in 
$H_n(M;\Z)$. Then we can study the functional $L(\Si)$ among embedded
hypersurfaces $\Si$
homologous to $\Si_0$.  In this case, $\Si\cup\Si_0$ is the boundary of a
domain with components
counted with multiplicity plus or minus one (according to the
orientations). We can then apply Stokes's theorem on each subdomain to
conclude that $\left|\int_{\Si}\Lambda\right|$ is bounded by Vol$\,M$
and $\int_{\Si_0}|\Lambda|$. The rest of the proof is the same as before.

\bigskip
We thank A. Strominger for discussions.
Research of E.W. has been supported in part by NSF Grant
PHY-95-13835 and the Caltech Discovery Fund.  Research of S.-T. Yau
supported in part by
NSF Grant DMS-9803347 and DOE Grant FG02-88ER25065.

\listrefs
\end